\documentclass[times,twocolumn,final]{elsarticle}
\usepackage{framed,multirow}

\usepackage{amssymb}
\usepackage{latexsym}

\usepackage{url}
\usepackage{xcolor}

\usepackage{hyperref}

\usepackage{graphicx}
\usepackage{tikz}
\usepackage{comment}
\usepackage{amsmath,amssymb} % define this before the line numbering.
\usepackage{color}
\usepackage{float}
\usepackage{marvosym}
\usepackage{paralist}
\usepackage{enumitem}
\usepackage{colortbl}
\usepackage{booktabs}
\usepackage{subcaption}
\usepackage{bbm}
\newcommand{\etal}{\textit{et al.}}

\definecolor{newcolor}{rgb}{.8,.349,.1}

\begin{document}

% \verso{K. Zhang \textit{et~al.}}

\begin{frontmatter}

\title{A Dataset for Deep Learning-based Bone Structure Analyses in Total Hip Arthroplasty}%
%\tnotetext[tnote1]{This is an example for title footnote coding.}

\author[1]{Kaidong Zhang}
\author[2]{Ziyang Gan}
\cortext[cor1]{Corresponding author: Ziyang Gan}
%   Tel.: +86-152-0983-2860}
\ead{richu@mail.ustc.edu.cn (Kaidong Zhang), gzyang84@163.com}
% \fntext[fn1]{This is author footnote for second author.}
\author[1]{Dong Liu}
%% Third author's email
%\ead{dongeliu@ustc.edu.cn}
\author[2]{Xifu Shang}

\address[1]{Department of Electronic Engineering and Information Science, University of Science and Technology of China, Hefei 230027, China}
\address[2]{Department of Orthopedics, The First Affiliated Hospital of the University of Science and Technology of China, Hefei 230001, China}

% \received{xxx}
% \finalform{xxx}
% \accepted{xxx}
% \availableonline{xxx}
%\communicated{S. Sarkar}

\begin{abstract}
Total hip arthroplasty (THA) is a widely used surgical procedure in orthopedics. For THA, it is of clinical significance to analyze the bone structure from the CT images, especially to observe the structure of the acetabulum and femoral head, before the surgical procedure. For such bone structure analyses, deep learning technologies are promising but require high-quality labeled data for the learning, while the data labeling is costly. We address this issue and propose an efficient data annotation pipeline for producing a deep learning-oriented dataset. Our pipeline consists of non-learning-based bone extraction (BE) and acetabulum and femoral head segmentation (AFS) and active-learning-based annotation refinement (AAR). For BE we use the classic graph-cut algorithm. For AFS we propose an improved algorithm, including femoral head boundary localization using first-order and second-order gradient regularization, line-based non-maximum suppression, and anatomy prior-based femoral head extraction. For AAR, we refine the algorithm-produced pseudo labels with the help of trained deep models: we measure the uncertainty based on the disagreement between the original pseudo labels and the deep model predictions, and then find out the samples with the largest uncertainty to ask for manual labeling. Using the proposed pipeline, we construct a large-scale bone structure analyses dataset from more than 300 clinical and diverse CT scans. We perform careful manual labeling for the test set of our data. We then benchmark multiple state-of-the-art deep learning-based methods of medical image segmentation using the training and test sets of our data. The extensive experimental results validate the efficacy of the proposed data annotation pipeline. The dataset, related codes and models will be publicly available at  \url{https://github.com/hitachinsk/THA}.
%%%%
\end{abstract}

% \begin{keyword}
%% Keywords
% \KWD Bone structure analysis\sep Dataset\sep Deep learning\sep Total hip arthroplasty
% \end{keyword}

\end{frontmatter}

%\linenumbers

\begin{figure*}[t]
\begin{center}
\includegraphics[width=\linewidth]{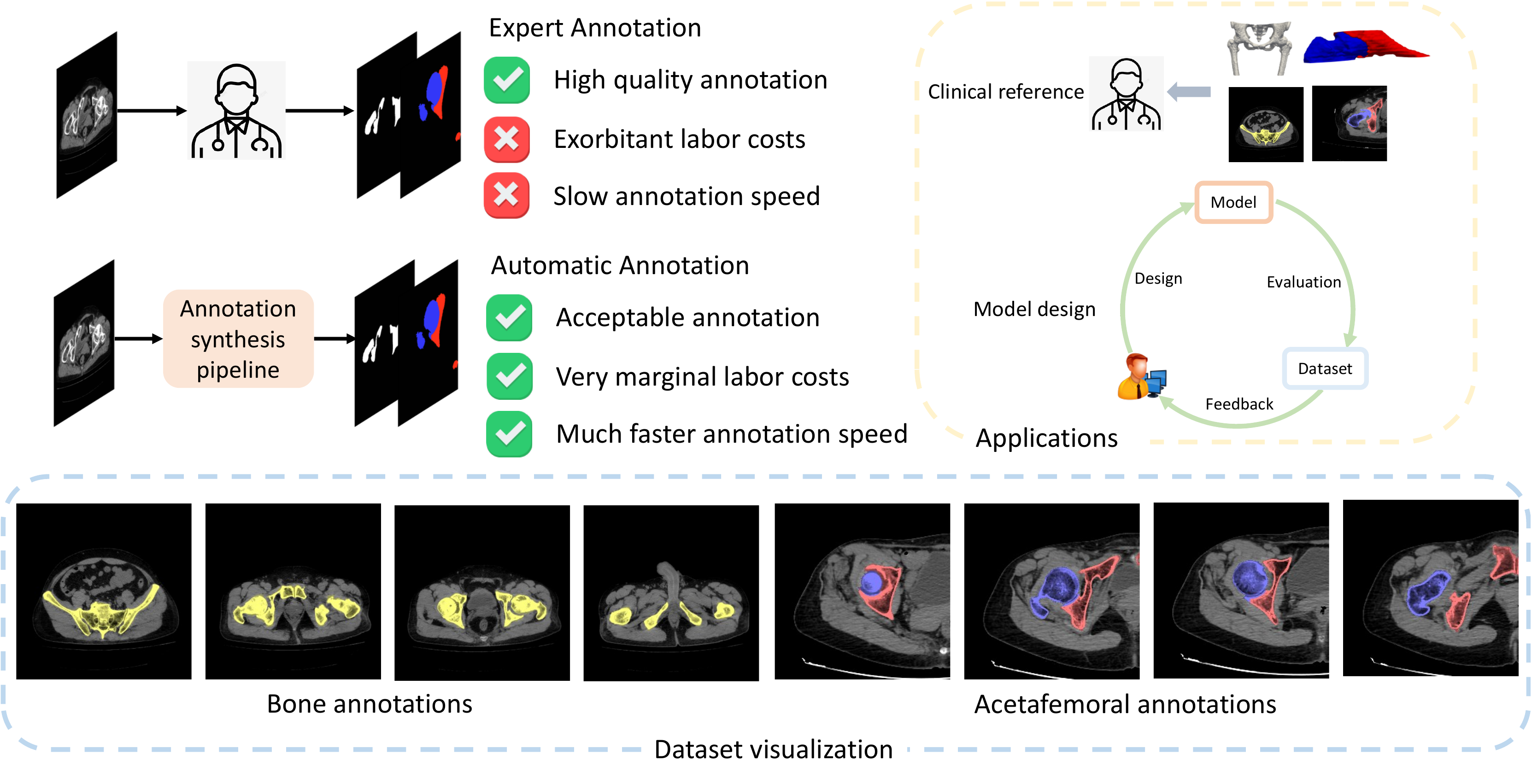}
\end{center}
   \caption{The comparisons of the advantages and disadvantages between expert annotation and our proposed annotation synthesis pipeline. We also present the application and the visualization of our dataset.}
\label{fig:teaser}
\end{figure*}

\section{Introduction}
\label{sec1}
% Please use \verb+elsarticle.cls+ for typesetting your paper.
% Additionally load the package \verb+medima.sty+ in the preamble using
% the following command: 
% \begin{verbatim} 
%   \usepackage{medima}
% \end{verbatim}

The surgical procedure known as Total Hip Arthroplasty (THA) has obtained widespread utilization in the management of debilitating conditions, including severe arthritis and fractures in the hip joint. This procedure involves the complete replacement of the damaged femoral head with a prosthetic implant, aiming to restore functionality and alleviate pain.  The prerequisite for THA lies in the precise analyses of the bone structure, which involves extracting bone from soft tissues and separating the acetabulum and femoral head. These processes hold significant importance for computer-assisted diagnosis and preoperative planning. 

Similar to other medical fields, it's necessary to build a large and diverse dataset to provide clinical references and empower deep learning for more precise bone structure segmentation in THA.  A trivial approach is to obtain multiple CT scans from professional organizations and annotate each scan manually. However, this manner is labor intensity and requires a huge amount of time and labor overhead from many professionally trained doctors. Although the dataset constructed in this manner is high quality, the huge overhead during construction hinders its scalability.

Fortunately, the specific anatomical structure of hip joint provides a shortcut to synthesize these annotations in an automatic way. With reasonable utility of such prior, we can obtain annotations with high fidelity and even comparable to human labelling. For those samples with severe arthropathy, although the prior anatomical structure is partially disrupted, we can still utilize the statistical differences among the samples to localize them, enabling targeted and refined labeling.

In this paper, we explore the annotation synthesis pipeline to generate semantic annotations for bone structure analyses of THA, and construct a large-scale dataset with our method to facilitate future research. Our annotation synthesis pipeline consists of three modules, including bone extraction (BE), acetabulum and femoral head segmentation (AFS) and active-learning-based annotation refinement (AAR). In BE, we exploit the difference of Hounsfield value between bone and soft tissues and utilize graph-cut~\cite{boykov2006graph} to extract the bone tissues. Similar to other anatomy-based methods~\cite{1263905,liu2016automatic}, AFS firstly estimates the boundary of femoral head in the initial slice, and then propagate the boundary to other slices that contain femoral head. But we make three improvements to achieve significantly higher annotation fidelity. First, we regularize the boundary of femoral head in the initial slice with the first-order and the second-order gradient (FSG) of the CT image simultaneously to filter out the false positive boundary points. Second, we adopt Line-based No-Max Suppression (L-NMS) to retain the outermost point along each normal direction in the boundary for more robust annotation generation. Finally, we leverage the anatomic circle of the femoral head to determine the best fitting boundary of the femoral head. Extensive experiments demonstrate the remarkable performance improvement of our innovations.

As for AAR, we refine the pseudo labels synthesized by our method with the help from the statistical prior captured by deep learning models. We firstly train a deep learning model with the annotations synthesized by our method. Then, we measure the uncertainty between the original annotations and the deep model predictions, and select a small number of samples with the highest uncertainty for manually reannotation. With the refined annotations, we observe a steady improvement of the newly trained model over the previous one trained on original annotations. Such phenomenon demonstrates the necessity and efficacy of active learning in annotation refinement. 

Based on our annotation synthesis pipeline, we construct a novel large-scale dataset for bone structure analyses in THA. Our dataset contributes the research community in the following aspects. (a) Large-Scale: There are 363 Computerized Tomography (CT) scans. We provide a mask that localize the bone tissues for each CT slice. As for the segmentation between acetabulum and femoral head, our dataset provides 251 volumes for training set, which contains 5297 slices. Moreover, we annotate 61 volumes, including 1003 slices manually to construct the testset for accurate measurement of the acetabulum and femoral head segmentation. (b) Diverse and clinical: We acquire the dataset in the real-world clinical settings. All the CT scans are from diverse patients with different kinds of abnormalities and diverse scanners. There also exists dramatic variance of the physical resolution in our dataset. Such diverse image and medical properties are beneficial for not only the clinical references but also more robust training of the deep learning models. We present the detailed pros and cons between annotation by experts and our proposed annotation synthesis pipeline, applications and dataset visualization in Fig.~\ref{fig:teaser}. In a nutshell, our method achieves fairly good annotation quality with very marginal labor and time costs. And our constructed dataset will facilitate the clinical reference of THA and empower the model development of bone analyses in hip joint.

We benchmark multiple open-sourced state-of-the-art deep learning methods with the training and test sets from our dataset. The contribution of this paper can be summarized as:
\begin{itemize}
    \item We propose a annotation generation pipeline to generate high quality annotations for bone analyses in hip joint automatically.
    \item We construct a large-scale dataset for bone structure analyses in THA with our proposed annotation generation pipeline.
    \item We benchmark our dataset on multiple open-sourced state-of-the-art medical image segmentation methods.
\end{itemize}

\section{Related work}
\subsection{Medical image segmentation}
Prior the prevalence of deep learning, medical image segmentation methods empirically adopt watershed~\cite{beucher1992watershed} or  graph-cut~\cite{boykov2006graph} as the basic segmentation tools and utilize explicit contour features~\cite{1194625} or Markov random field~\cite{held1997markov} as guidance for more accurate segmentation results. Nowadays, with the increase in computing power, a proliferation of deep-learning based approaches emerge and achieve unprecedented performance in medical image segmentation. These methods always follow U-Net~\cite{ronneberger2015u} and design more advanced structure with convolutional neural network, including Res-UNet~\cite{xiao2018weighted}, Dense-UNet~\cite{li2018h}, U-Net++~\cite{zhou2019unetplusplus} and 3D-Unet~\cite{cciccek20163d}. Recently, with the ubiquitous application of Transformer~\cite{vaswani2017attention} in computer vision, some research works explore the transformer-based structures in medical image segmentation. Representative works are TransUnet~\cite{chen2021transunet}, SwinUnet~\cite{swinunet}, MISSFormer~\cite{9994763}, DAE-Former~\cite{azad2023daeformer}, etc. Extracting bones or segmentation between acetabulum and femoral head can be regarded as a subset of medical image segmentation with certain classes.

\begin{figure*}[t]
\begin{center}
\includegraphics[width=\linewidth]{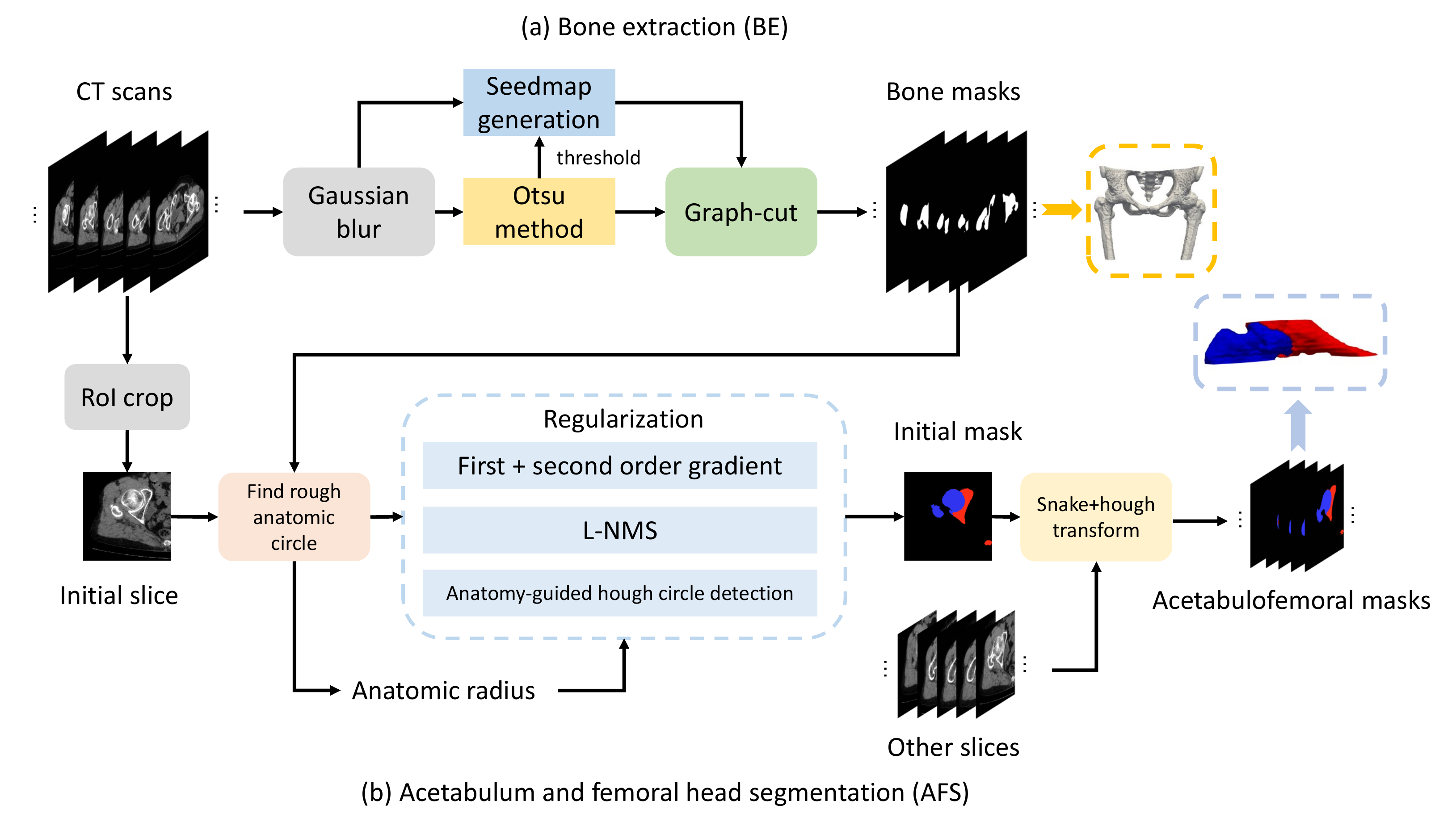}
\end{center}
   \caption{The illustration of BE (a) and AFS (b) in the annotation synthesis pipeline. With BE, we can extract bone structures from the CT scans. AFS is designed for the generation of acetabulofemoral masks, which can segment the bone structure between acetabulum and femoral head.}
\label{fig:pipeline}
\end{figure*}

\subsection{Acetabulum and femoral head segmentation}
There are multiple segmentation methods designed for segmentation between acetabulum and femoral head specifically. These methods exploit the specific structure in the hip joint and can be roughly divided into four types. Thresholding-based methods~\cite{beucher1992watershed,boykov2006graph,5872823} firstly determine the patient-specific optimal threshold value for reliably separating the bone structure in the hip joint. Atlas based methods~\cite{CHU2015173,tabrizi2015acetabular}  construct and map the bone-normalized probabilistic atlas for accurate segmentation. Shape-model based methods~\cite{ALMEIDA20161474,ben2014fully,lindner2013fully,10.1007/978-3-642-40763-5_24,10.1007/978-3-642-04271-3_98} construct Statistical Shape Model (SSM) to complete segmentation of the pelvis and femur accurately. Anatomy-based methods~\cite{1263905,cheng2013automatic,liu2016automatic,liu2014improve} leverage the specific structure in the hip joint to perform accurate segmentation. Compared with previous types of methods, anatomy-based methods do not need to annotate the labels to construct atlas/SSM or calculate the threshold heuristically. Therefore, we develop the annotation synthesis pipeline based on the anatomy-based method for accurate automatic annotation with marginal labor and time costs.

Currently, some researchers also explore the feasibility of CNN in the segmentation between femoral head and acetabulum~\cite{deniz2018segmentation,chen2017three,wu2022utility}. But these methods needs a huge amount of manually annotated labels for accurate segmentation. Moreover, these methods did not release the dataset, which reflects the necessity of proposing a professional dataset for accurate segmentation and bone analyses in the hip joint.

\subsection{Active learning}
Active learning~\cite{settles2009active} aims to find out the most valuable data samples in the dataset to improve the annotation quality. To accomplish active learning, we can measure the uncertainty produced from the models~\cite{kendall2017uncertainties,gal2017deep,beluch2018power}, the representativeness constructed by the model~\cite{sener2017active,ash2019deep}, the training effects based on current annotated results~\cite{NIPS2007_a1519de5,houlsby2011bayesian,toneva2018empirical} and the hybrid of the methods above~\cite{yang2017,zhdanov2019}. In this work, we measure the uncertainty between the prediction from the deep learning model and the original annotations to determine the subset of samples for manual reannotation. Our method only annotate a very marginal subset of the training labels and obtain stable improvement in the segmentation accuracy from the trained deep learning models, which demonstrates the efficacy of utilization of active learning in the label refinement of our dataset.

\section{Annotation synthesis pipeline}
\subsection{Bone extraction}~\label{bse}
To locate the bone structure in the hip joint, we firstly extract the bone structure from the CT scans. Bone extraction (BE) is based on the graph-cut segmentation framework~\cite{boykov2006graph}. Given a CT slice $\boldsymbol{S}$, our goal is to obtain the segmentation mask that delineates the bone tissues $\boldsymbol{M}_B$, which is a binary mask, and ``1" represents the regions that contain bone tissues. The graph-cut algorithm can be described as,
\begin{equation}
    \zeta: \boldsymbol{S} \rightarrow \boldsymbol{M_B}.
\end{equation}
Graph-cut is an energy-based method and requires two components to define the cost function: the per-pixel term $R_p(\boldsymbol{M}_B(p))$ and the boundary term $B(p, q)$, where $p$ and $q$ represent the position of the pixels, and $\boldsymbol{M}_B(p)$ is the binary labelling in position $p$. Therefore, the energy function can be defined as,
\begin{equation}
    E(M_B) = \sum_{p\in S} R_p(\boldsymbol{M}_B(p))+\lambda\sum_{(p,q)\in \mathcal{N}}\delta(\boldsymbol{M}_B(p), \boldsymbol{M}_B(q))B(p,q).
\end{equation}
Where $(p,q) \in \mathcal{N}$ indicates $p$ and $q$ should be the neighborhood of the CT image. $\delta$ is the Kronecker delta and $\lambda$ aims to strike a balance between he per-pixel term and the boundary term. 

The pipeline of BE is illustrated in Fig.~\ref{fig:pipeline}(a). We firstly adopt Gaussian blur to reduce the interference from the noise in CT scans, and determine the seed map with the threshold calculated by Otsu threshold method~\cite{4310076}. With the initialized seed map, we perform graph-cut algorithm for each CT scans.

\subsection{Acetabulum and femoral head segmentation}

After we extract the bone structure in each CT scans, we segment the acetabulum and the femoral head in the CT volumes that contain these two structures simultaneously. Our method adopts the anatomy prior between acetabulum and femoral head and  follows the segmentation procedure that segments the initial slice first and propagate the segmentation mask from initial slice to the other slices in the volume, which is a common practice in anatomy-based acetabulum and femoral head segmentation~\cite{1263905, liu2016automatic}. We illustrate the pipeline of AFS in Fig.~\ref{fig:pipeline}(b).

\begin{figure}[t]
\begin{center}
\includegraphics[width=1.0\linewidth]{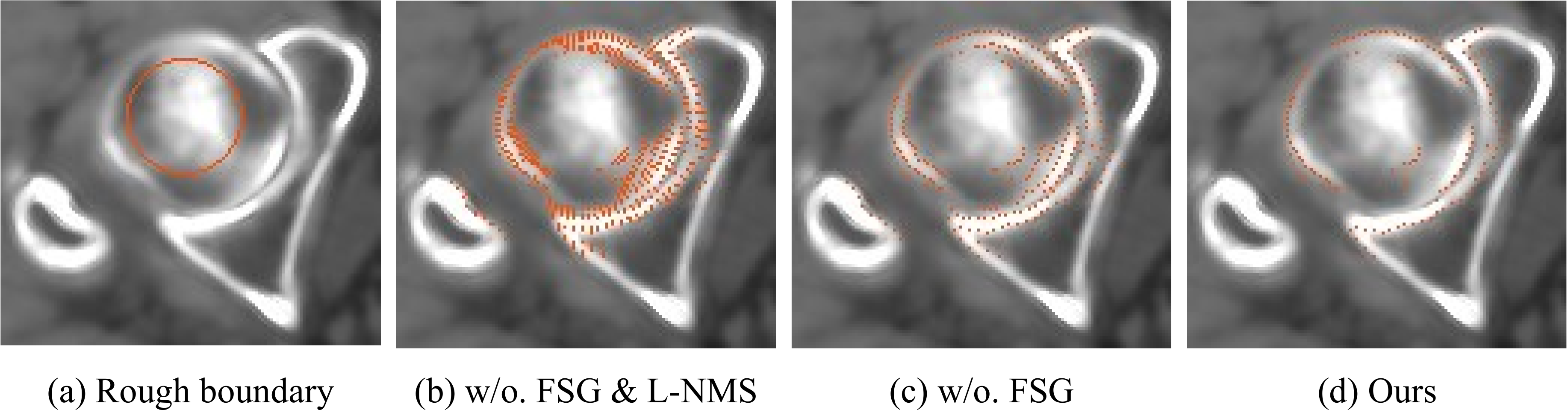}
\end{center}
   \caption{The illustration of the rough boundary and the variance of refined boundary synthesized with or without our proposed regularization method. The red points are the detected refined boundary points. FSG: first-order and second-order regularization. L-NMS: line-based no-max suppression.}
\label{fig:gen_mask_vis}
\end{figure}

We first extract the region of interest from each CT slice. Given $\boldsymbol{M}_B$, we crop a rectangle around its centroid whose height and width are half of the original CT slice. Next, we synthesize the boundary of femoral head in the initial slice. We define the initial slice as the CT slice in which the greater trochanter and femoral head are separated from each other. We approximate the boundary with a rough anatomic circle, and this circle is determined by three feature points. We denote the bone mask of the initial slice generated from BE as $\boldsymbol{Ms}_{B}$. We calculate the convex hull and the centroid $p_c$ of $\boldsymbol{Ms}_{B}$, and then cast rays in the up and right directions from $p_c$. The intersections of the rays and the convex hull are denoted as $p_u$ and $p_r$. Finally, we search the farthest point from the greater trochanter to $p_c$ and denote this feature point as $p_m$. We calculate the circle with the feature points $p_u$, $p_m$ and the midpoint between $p_c$ and $p_r$ to obtain the rough boundary $\boldsymbol{R}_b$ of the femoral head. And we denote the radius $r$ of $\boldsymbol{R}_b$ as the anatomy radius of the femoral head. Since some of the boundary points in $\boldsymbol{R}_b$ may be outside femoral head, and these boundary points are hard to be refined, we determine the radius of the rough boundary to be $1/1.7$ of $r$ to make sure all the boundary points inside the femoral head. We illustrate the detected rough boundary in Fig.~\ref{fig:gen_mask_vis} (a).

The rough boundary can only describe the rough structure of the femoral head, which is far not enough for a valid annotation of the dataset. Therefore, the refinement of the rough boundary is crucial in constructing a dataset with high fidelity. Liu \etal~\cite{liu2016automatic} utilizes Hessian filter~\cite{10.1007/BFb0056195} to enlarge the contrast in the narrowness between acetabulum and femoral head and adopts the second order gradient to select the candidate points of the refined boundary,
\begin{equation}
\label{eq1}
\boldsymbol{F}_b(p+1)=\left\{
\begin{aligned}
&1 , & \Delta \boldsymbol{S}(p+1)- \Delta \boldsymbol{S}(p)>0, \\
&0 , & \Delta \boldsymbol{S}(p+1)-\Delta \boldsymbol{S}(p)\le 0.
\end{aligned}
\right.
\end{equation}
Where $p$ is the point along the normal of the target direction, and the boundary refinement process gradually pushes the rough boundary to the cortex of the femoral head. However, as shown in Fig.~\ref{fig:gen_mask_vis}(b), such boundary refinement strategy leads to multiple false candidate boundary points along the normal direction and in the spongy bones of the femoral head, which greatly deteriorates the accuracy of the quality of the detected femoral head. 

The philosophy behind formula~\ref{eq1} is that the rich calcium in cortex of the boundary of femoral head and the surrounding soft tissues forms strong contrast, which can be reflected in the dramatic variance of the gradient. However, these regions are not the only one that can reflect such great contrast. Due to the existence of spongy tissues or degeneration in the bones, these regions can also form strong contrast with the other bone tissues, which leads to false detected candidate boundaries. Moreover, the bone structure is continuous, which means if point $p$ is selected as the candidate boundary, the probability of selecting point $p+1$ as the candidate boundary will increase.  And these continuous points will form the ``lines" along each normal direction, greatly undermines the accuracy of the refined femoral head boundary.

Therefore, we design three strategies to further regularize the boundary refinement procedure for fitting the boundaries of the femoral head more accurately. First, we propose the synergy of first order gradient and second order gradient to determine the refined candidate boundary, which can be formulated as
\begin{equation}
\label{eq2}
\boldsymbol{F}_b(p+1)=\left\{
\begin{aligned}
&1 , & \Delta \boldsymbol{S}(p+1)- \Delta \boldsymbol{S}(p)>0~\&~\Delta \boldsymbol{S}(p+1)>0, \\
&0 , & else .
\end{aligned}
\right.
\end{equation}

Since the cortex is located on the inner side of the soft tissues and cortex accumulated more calcium than other regions, the first-order gradient of the boundary of femoral head must be greater than zero. Therefore, the first-order gradient can filter out the false candidate femoral head boundary points constructed by the dramatic variance between spongy bones and other bone tissues, as shown in Fig.~\ref{fig:gen_mask_vis} (c) and (d).

As for the continuous lines along the normal direction, we utilize the line-based non-max suppression (L-NMS) to retain the outermost boundary points, which are more likely to be the boundary of the femoral head. This operation can be described as
\begin{equation}
    \boldsymbol{F}_b(p)=0~if~\boldsymbol{F}_b(p+1) \times \boldsymbol{F}_b(p)=1  .
\end{equation}
As illustrated in Fig.~\ref{fig:gen_mask_vis} (b) and (c), L-NMS can filter out the continuous points in the normal direction and the first order gradient is crucial for the elimination of the points that are selected due to the variance between spongy bones and the other bone tissues, which improve the quality of the synthesized annotations significantly.

Finally, we apply Hough transform to fit the circle formed by the refined boundary and utilize snake algorithm~\cite{kass1988snakes} to propagate the boundaries to other slices in the same volume iteratively. Hough transform may return multiple detected circles and their corresponding confidence, and which circle to adopt as the final boundary of the femoral head is worth to investigate. A naive approach is to adopt the circle with the largest confidence. However, such approach ignores the anatomy prior of the femoral head, and the radius of the detected boundary may vary a lot from that of the rough boundary. Therefore, we select the circle that owes the closest radius with the anatomy radius  $r$ as the detected final boundary of the femoral head.

\begin{figure}[t]
\begin{center}
\includegraphics[width=1.0\linewidth]{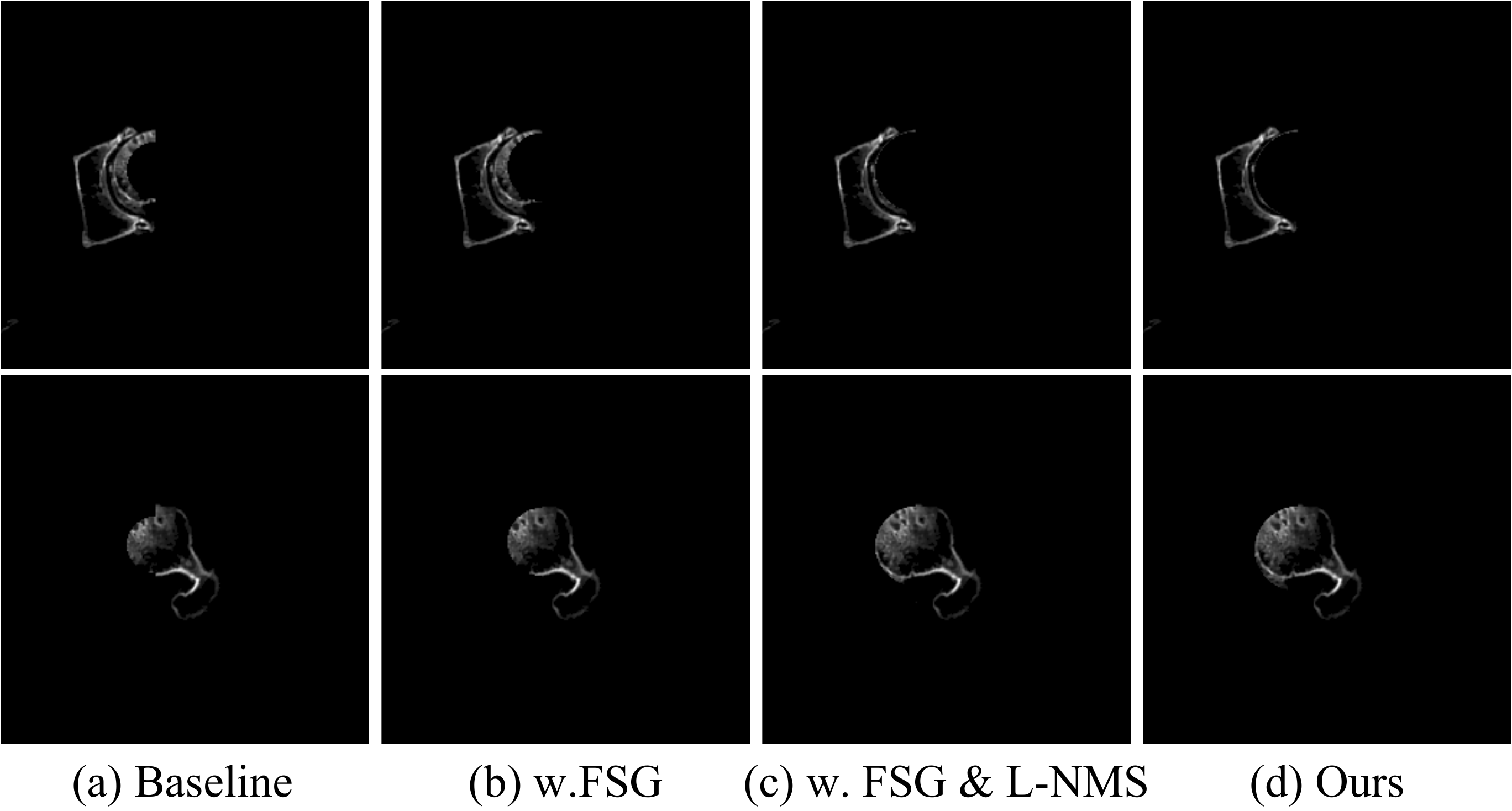}
\end{center}
   \caption{The results of the extracted acetabulum (the first row) and the femoral head (the second row) with the synthesized acetabulofemoral masks. Baseline: The method without FSG, L-NMS and AHS. Ours: our method with all the regularization methods.}
\label{fig:SAF_compare}
\end{figure}

Starting from the basic solution without our proposed regularization (We denote this method as ``Baseline"), we gradually add our innovations and extract the femoral head and the acetabulum with the synthesized acetabulofemoral masks for comprehensive visualization. The results in Fig.~\ref{fig:SAF_compare} validate our innovations. With our innovations, we can capture more accurate boundary between acetabulum and femoral head, which is crucial for construction of a high quality deep-learning based bone structure analyses dataset for THA.

\begin{figure}[t]
\begin{center}
\includegraphics[width=1.0\linewidth]{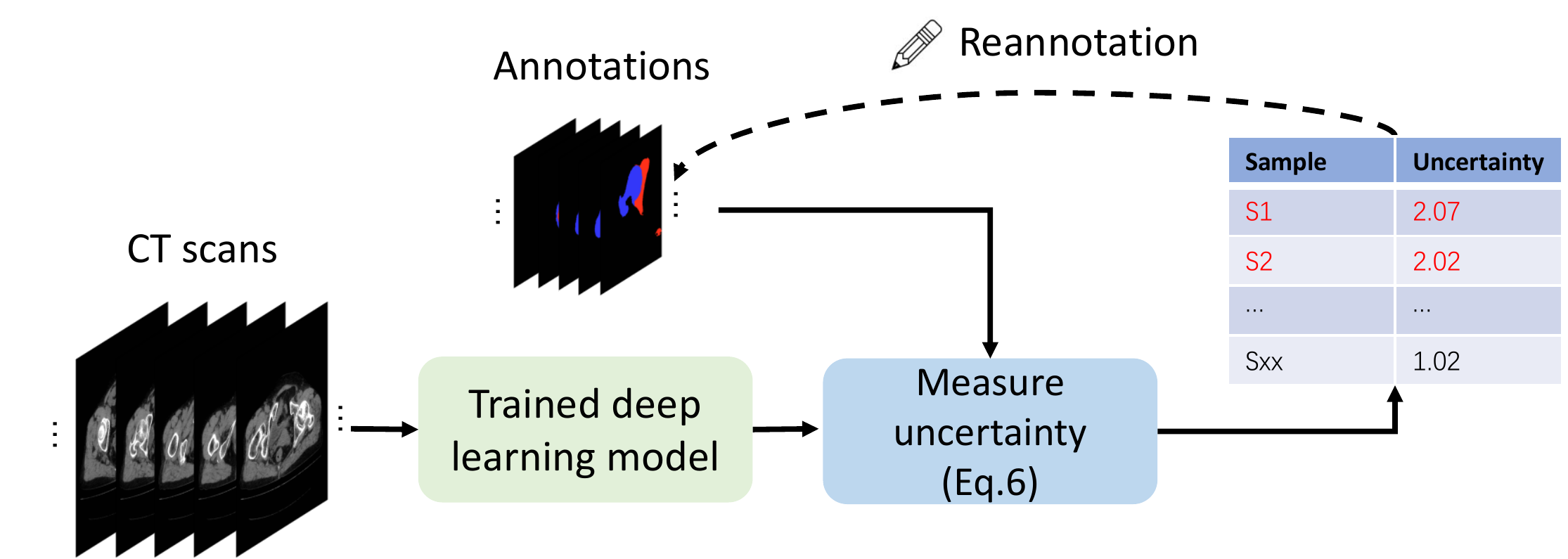}
\end{center}
   \caption{The pipeline of active learning based annotation refinement (AAR).}
\label{fig:AAR}
\end{figure}

\subsection{Active learning based annotation refinement}~\label{active}
The segmentation between acetabulum and femoral head utilizes the anatomy prior to synthesize the annotations, which leads to high quality annotation generation for most of the CT scans. However, when the degeneration is severe or the intrinsic bone structure of the patient is abnormal, the annotation quality may degrade severely due to the lack of anatomy prior. A reasonable strategy for annotation improvement is to localize and reannotate these samples manually. 

We propose to leverage the statistical prior from deep learning models to improve the annotation quality based on active learning. We firstly train a deep learning model $F_{\theta}$ with the annotated dataset. The statistical knowledge and the structural prior~\cite{ulyanov2018deep} of the deep learning models influence the segmentation results. For each sample from the training set, if the prediction from the trained deep learning model $\boldsymbol{\hat{M}}_s$ is largely different from the original annotation $\boldsymbol{M}_s$, the knowledge in the trained model may deviate from the general anatomy prior, which means this sample needs to be reannotated. Otherwise, the annotation of the sample fits the anatomy prior, and the reward of reannotation is marginal.

\begin{figure}[t]
\begin{center}
\includegraphics[width=0.9\linewidth]{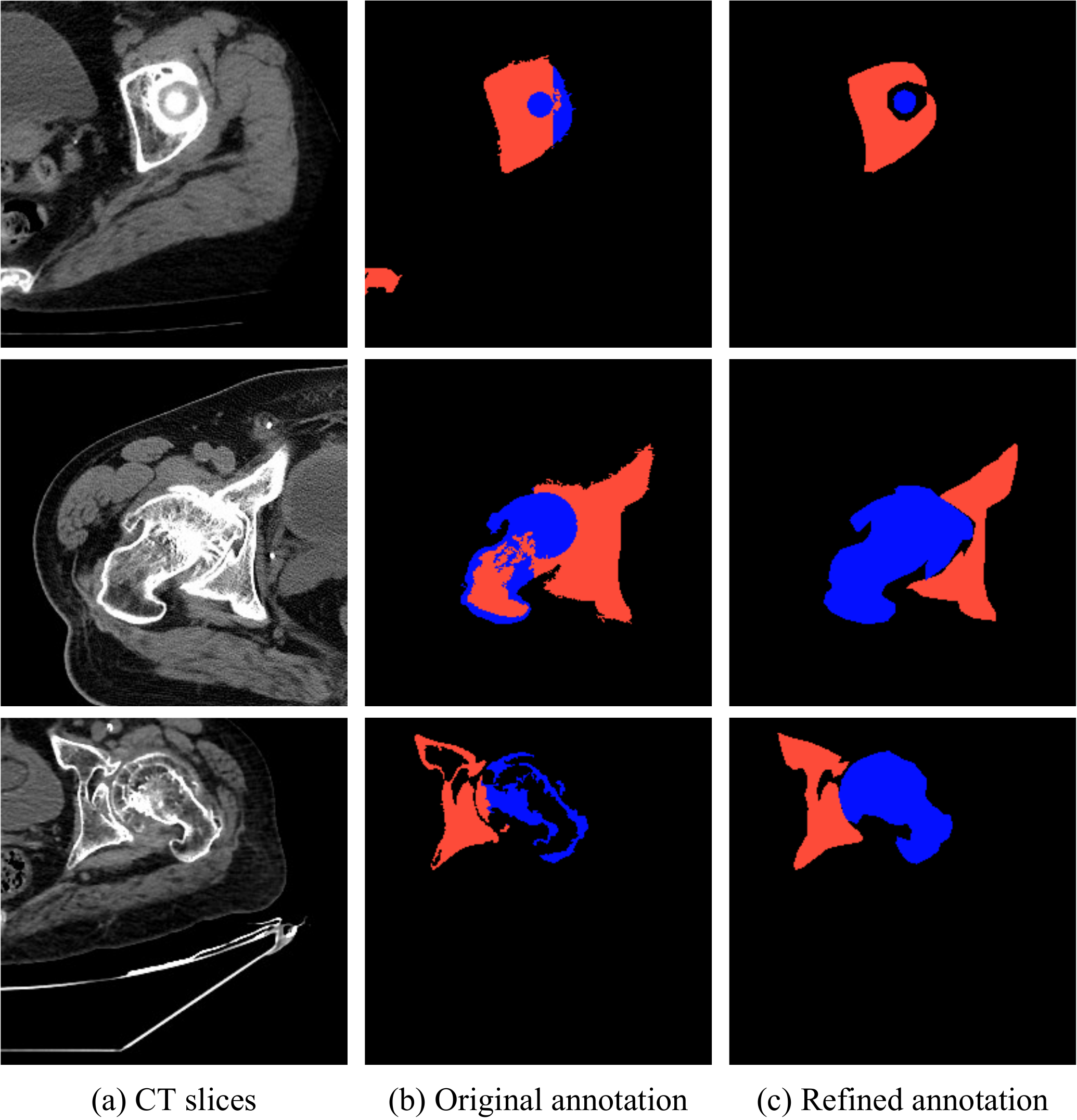}
\end{center}
   \caption{The visualization of the localized CT slices, original labels and the refined labels with AAR.}
\label{fig:AAR_vis}
\end{figure}

We term the ``disagreement" between $\boldsymbol{\hat{M}}_s$ and $\boldsymbol{M}_s$ as  the uncertainty of the sample, and propose to quantify the uncertainty $\mathcal{U_s}$ with the inverse of the dice score between $\boldsymbol{\hat{M}}_s$ and $\boldsymbol{M}_s$, which can be described as,
\begin{equation}
    \mathcal{U_s}=\frac{1}{Dice(\boldsymbol{\hat{M}}_s, \boldsymbol{M_s})+\epsilon}=\frac{|\boldsymbol{\hat{M}}_s+\boldsymbol{M}_s|}{\boldsymbol{\hat{M}}_s \cap \boldsymbol{M}_s+\epsilon}.
\end{equation}
Where $\epsilon$ is a very small number to prevent the zero division error.

The pipeline of AAR is illustrated in Fig.~\ref{fig:AAR}. We measure the uncertainty of all the samples in the training set, and select K samples with the highest uncertainty for manual reannotation. Since manual annotating procedure needs much higher cost compared with the annotation synthesis pipeline, we constrain the number K. In practice, we set K to 200, which means the samples for annotating again only occupies about 3.8\% of the total samples. Although we only annotate a very small amount of samples, we can still observe a steady improvement of the generalization ability of the deep learning  models trained on the dataset with AAR refinement, which demonstrate the necessity and effectiveness of AAR. We show some examples localized and reannotated by AAR in Fig.~\ref{fig:AAR_vis}. The samples selected by AAR often exhibit severe pathology at the hip joint or at the edge of these regions, and the anatomy-based annotation synthesis pipeline cannot achieves excellent results on these samples because they always do not obey the anatomy prior. Therefore, utilizing AAR for the localization and reannotation of these samples can effectively improve the quality of the annotation of these samples.

\section{Dataset}

\begin{table}[t!]
\centering
\caption{The details of CT scanners and the number of volumes and slices generated from them. All the vendors of the CT scanners are GE Medical System except NeuViz 128.}
\footnotesize
% \resizebox{\textwidth}{!}{
\begin{tabular}{ccc}
\toprule
Scanner & Volumes & Slices\\
\midrule
CT 99 & 69 & 3348 \\
NeuViz 128 & 248 & 26123  \\
CT 680 & 42 & 4140 \\
CT 750 & 4 & 499 \\
\bottomrule
\end{tabular}
% }
\label{scanner}
\end{table}

Based on our annotation synthesis pipeline for bone analyses in THA, we construct a large-scale dataset to facilitate the research in this field. All the annotations in the training set are synthesized automatically, therefore the annotation overhead of our dataset can get greatly reduced.

We follow the standard clinical acquisition protocols and acquire CT volumes from the first affiliated hospital of the university of science and technology of China. We acquire the CT slices in axial direction, and collect the CT volumes from 128 patients who are diagnosed with hip joint degeneration and receive THA surgery. We de-identified the dataset for privacy concern and the data format of all the CT volumes is DICOM image series. All the CT volumes are scanned by one of the four machines to reduce the data bias induced by a single CT scanner. The details of the CT scanners and the number of CT volumes and slices obtained from them are shown in Tab.~\ref{scanner}.

\begin{figure}[t]
    \centering
    \subcaptionbox{spatial resolution \label{spa_res}}[0.49\linewidth]
    {
        \includegraphics[width=1\linewidth]{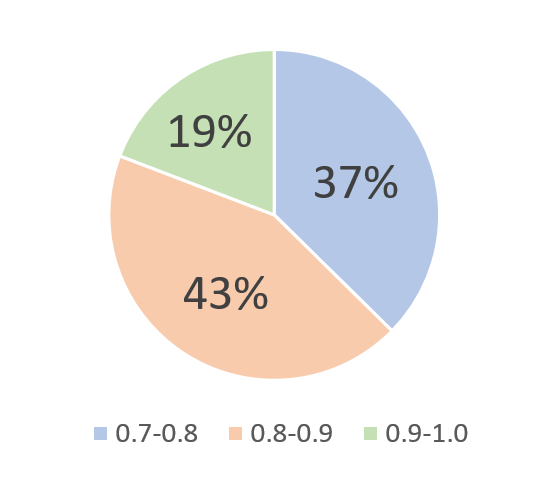}
    }
    \subcaptionbox{axial resolution \label{axial_res}}[0.49\linewidth]
    {
        \includegraphics[width=1\linewidth]{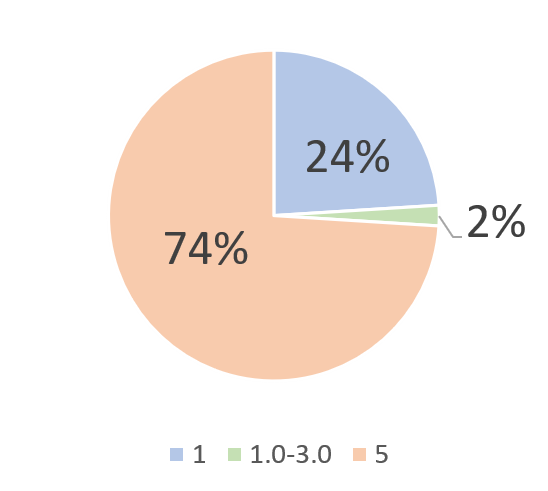}
    }
    % \vspace{-0.3cm}
    % \renewcommand{\captionfont}{\normalsize} 
    \caption{The distribution of spatial and axial resolution of the dataset.}
    \label{fig:resolution}
    % \vspace{-0.4cm}
\end{figure}

Each CT slice contains 512$\times$512 pixels. The spatial resolution varies from 0.7 to 1.0 and the range of axial resolution is from  1.0 to 5.0. The percentage of CT volumes included in the total CT volumes  for different spatial and axial resolutions are shown in Fig.~\ref{fig:resolution}.

\begin{table}[t!]
\centering
\caption{The number of the annotated volumes and slices for the segmentation between acetabulum and femoral head.}
\footnotesize
% \resizebox{\textwidth}{!}{
\begin{tabular}{ccc}
\toprule
Data split & Volumes & Slices\\
\midrule
Training & 251 & 5297 \\
Test & 61 & 1003  \\
\bottomrule
\end{tabular}
% }
\label{ace_fem}
\end{table}

Each CT slice is occupied with a mask that describes the bone structure. As for the segmentation between acetabulum and femoral head, we extract the slices that contain both acetabulum and femoral head in each CT volume. All the annotations are synthesized by our proposed annotation synthesis pipeline. Moreover, we select the CT scans that the annotation synthesis pipeline fails to generate accurate acetabulofemoral annotations and annotate these scans manually to construct the testset for quantitative measurement of the segmentation between acetabulum and femoral head. All the manual annotations are corrected and confirmed by the experts from the first affiliated hospital of the university of science and technology of China. The numbers of the volumes and slices in the training and test set for segmentation between acetabulum and femoral head are shown in Tab.~\ref{ace_fem}.

\begin{table*}[t!]
\fontsize{8}{9}\selectfont
\centering
\caption{Quantitative benchmark on our proposed dataset for the segmentation between femoral head and acetabulum}
% \footnotesize
% \resizebox{\textwidth}{!}{
\setlength{\tabcolsep}{1mm}{
\begin{tabular}{cccccccc}
\toprule
\multirow{2}{*}{Type} & \multirow{2}{*}{Method} & \multicolumn{2}{c}{Average} & \multicolumn{2}{c}{Acetabulum} & \multicolumn{2}{c}{Femoral head}\\ \cmidrule(l){3-4} \cmidrule(l){5-6} \cmidrule(l){7-8}
& & DSC$\uparrow$ & HD$\downarrow$ & DSC$\uparrow$ & HD$\downarrow$ & DSC$\uparrow$ & HD$\downarrow$ \\ \midrule
\multirow{2}{*}{Convolution-based} & U-Net~\cite{ronneberger2015u} & 87.79 & 8.05 & 86.52 & 11.66 & 89.05 & 4.44 \\
& Att-UNet~\cite{oktayattention} & 86.74 & 7.84 & 86.19 & 11.20 & 87.29 & 4.47 \\ \midrule
\multirow{4}{*}{Transformer-based} & TransUnet~\cite{chen2021transunet} & 90.15 & 7.45 & 87.75 & 11.12 & 92.55 & 3.79 \\
& SwinUnet~\cite{swinunet} & 90.43 & 7.41 & 87.93 & 11.20 & 92.94 & 3.63 \\
& MissFormer~\cite{9994763} & 90.74 & 7.19 & 88.51 & 10.63 & 92.97 & 3.74 \\
& DAE-Former~\cite{azad2023daeformer} & 90.67 & 7.20 & 88.37 & 10.71 & 92.96 & 3.70 \\ \midrule
PEFT-based & SAMed & 89.08 & 8.12 & 86.46 & 12.24 & 91.69 & 4.00 \\
\bottomrule
\end{tabular}}
% }
\label{benchmark}
\end{table*}

\section{Experiments}
\subsection{Settings} We mainly benchmark our dataset on acetabulum and femoral head segmentation task. Due to the narrowness between pelvis and femur in the joint space, this task is more difficult and representative. Therefore, as shown in Tab.~\ref{ace_fem}, we adopt the training set to train the models, including 251 volumes and 5297 slices and the test set to measure the performance between different methods. We adopt the average dice score (DSC) and the average Hausdorff distance (HD) as the objective metrics.

Before training a deep learning model with our dataset, we preprocess the dataset for successful training. We clip the data range of each CT scans from -125 to 275, and normalize each CT scans to 0$\sim$1. Since the bone tissues are continuous in both acetabulum and femoral head, we fill the holes in the synthesized annotations. Moreover, we extract the region of interest from each CT scan in advance. Therefore, the input resolution of each CT slice is 256$\times$256. 

\subsection{Quantitative benchmark} We provide a comprehensive quantitative benchmark on multiple open-sourced medical image segmentation methods on our dataset, including convolution-based U-Net~\cite{ronneberger2015u} and Att-UNet~\cite{oktayattention}; transformer-based TransUnet~\cite{chen2021transunet}, SwinUnet~\cite{swinunet}, MissFormer~\cite{9994763}, DAE-Former~\cite{azad2023daeformer} and a novel parameter efficient fine-tuning (PEFT) method SAMed~\cite{samed}.

We report the benchmark in Tab.~\ref{benchmark}. Our dataset can make all of the methods converge, which reflects the annotations synthesized by our method show a uniform and stable pattern for the deep learning models.  All the methods achieve relatively powerful performance on our dataset, and the transformer-based methods and the PEFT-based method achieve higher performance on both dice score and Hausdorff distance, which means the more advanced network structure and training technology could lead to more accurate segmentation ability.

\subsection{Qualitative comparisons} 

\begin{figure}[t]
\begin{center}
\includegraphics[width=1.0\linewidth]{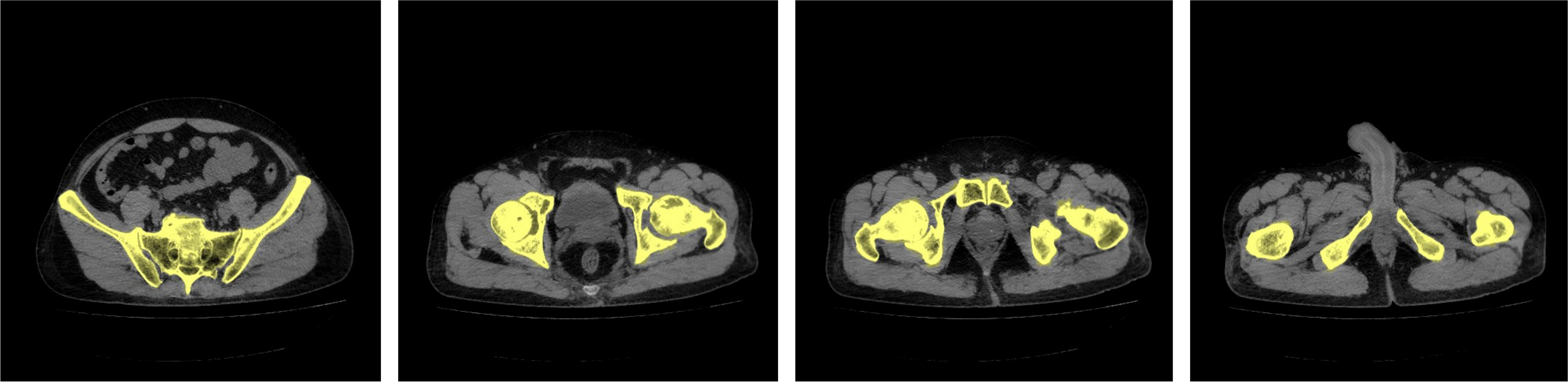}
\end{center}
   \caption{The samples of the bone structure extracted by BE. We color the bone structure in the images yellow.}
\label{fig:bone_2d}
\end{figure}

\begin{figure}[t]
\begin{center}
\includegraphics[width=1.0\linewidth]{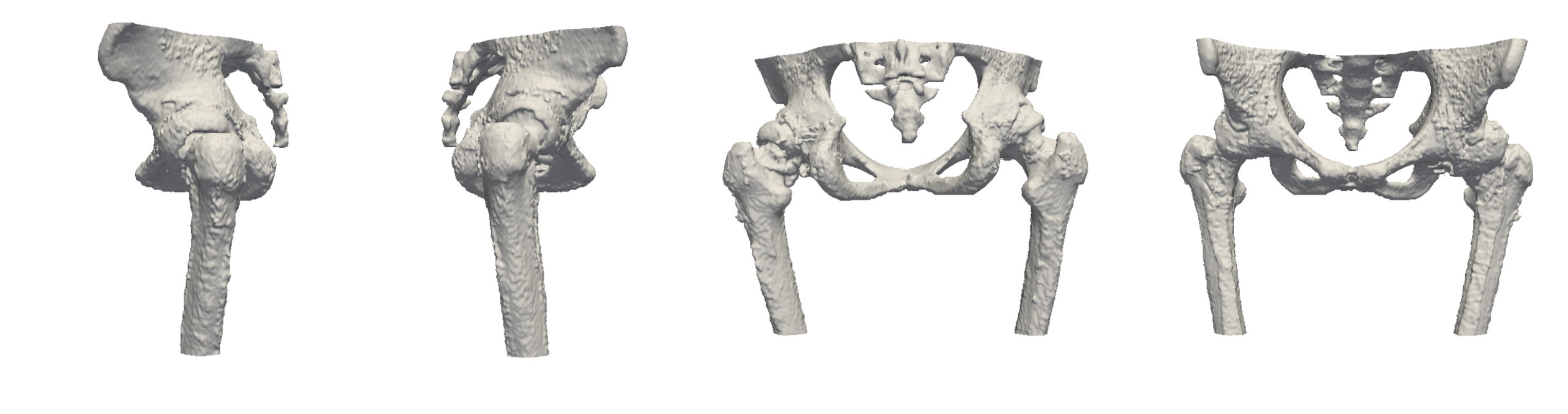}
\end{center}
   \caption{The illustration of the 3D bone structure extracted by BE. We present the bone structure from different views for comprehensive visualization.}
\label{fig:bone_3d}
\end{figure}

\subsubsection{Bone extraction} First, we present the CT slices of the bone structure extracted by BE in Fig.~\ref{fig:bone_2d}. Although BE is simple and only contains the graph-cut algorithm and the Otsu based seedmap generation, the bone structure extracted by BE is with high fidelity and maintains the bone structure well. Therefore, BE is capable of extracting the entire bone structure. Fig.~\ref{fig:bone_3d} depicts the 3D illustration of the bone structure for comprehensive visualization. We render the extracted 3D structure from different views. In general, BE not only extracts the bone structure of each slice accurately, but also maintains the overall integrity of the bone structure from 3D perspective, which benefits the sequential synthesizing of the acetabulofemoral masks for the segmentation between acetabulum and femoral head.

\begin{figure}[t]
\begin{center}
\includegraphics[width=1.0\linewidth]{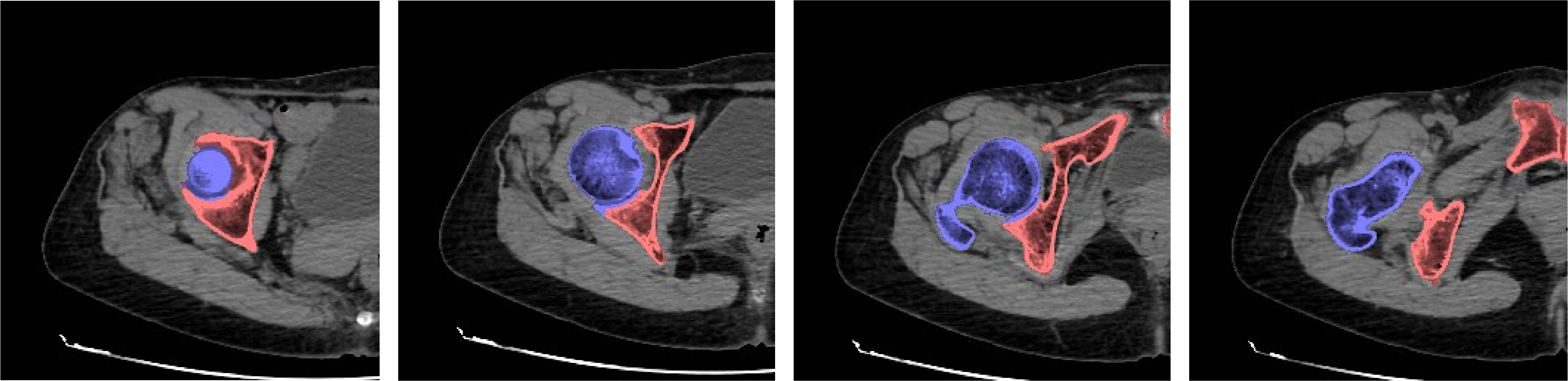}
\end{center}
   \caption{The acetabulofemoral masks synthesized by our proposed AFS and AAR. We color acetabulum red, and femoral head blue.}
\label{fig:train_2d}
\end{figure}

\begin{figure}[t]
\begin{center}
\includegraphics[width=1.0\linewidth]{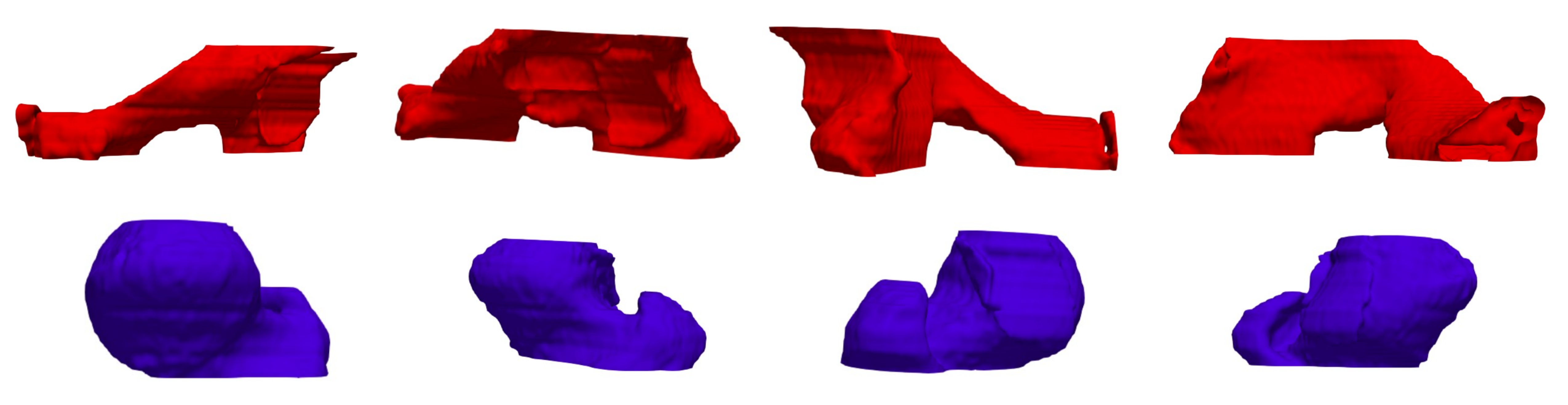}
\end{center}
   \caption{The visualization of 3D bone structure synthesized by AFS and AAR. We visualize both acetabulum and femoral head in four different views. The top row shows acetabulum (red) and the bottom presents the femoral head (blue).}
\label{fig:train_3d}
\end{figure}

\begin{figure}[t]
\begin{center}
\includegraphics[width=1.0\linewidth]{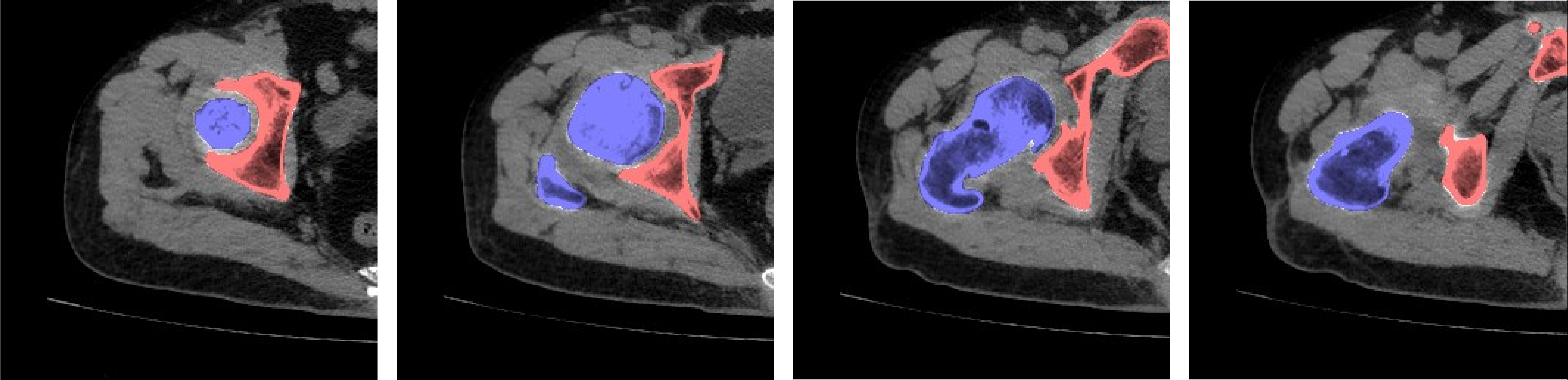}
\end{center}
   \caption{The manually annotated segmentation masks for evaluation of the segmentation between acetabulum and femoral head. Blue regions: acetabulum. Red regions: femoral head.}
\label{fig:test_2d}
\end{figure}

\begin{figure}[t]
\begin{center}
\includegraphics[width=1.0\linewidth]{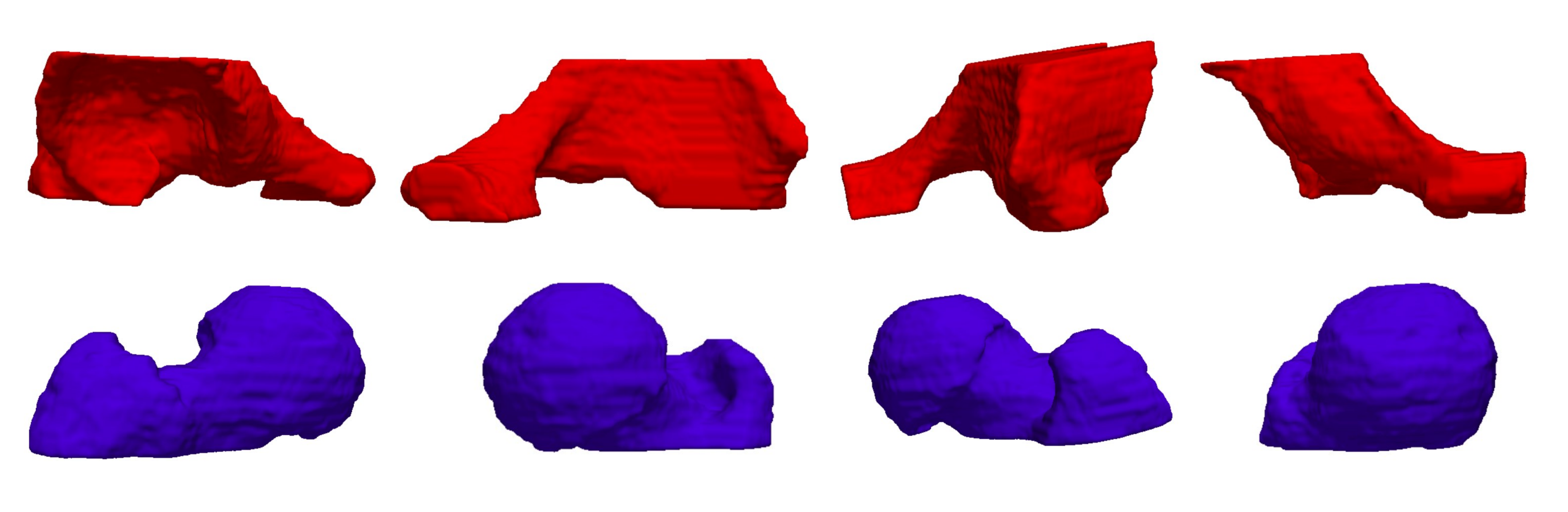}
\end{center}
   \caption{The 3D structure constructed by each manually annotated slices. We render the 3D structure from four different views. Top row: acetabulum (red); bottom row: femoral head (blue).}
\label{fig:test_3d}
\end{figure}

\begin{figure*}[t]
\begin{center}
\includegraphics[width=1.0\linewidth]{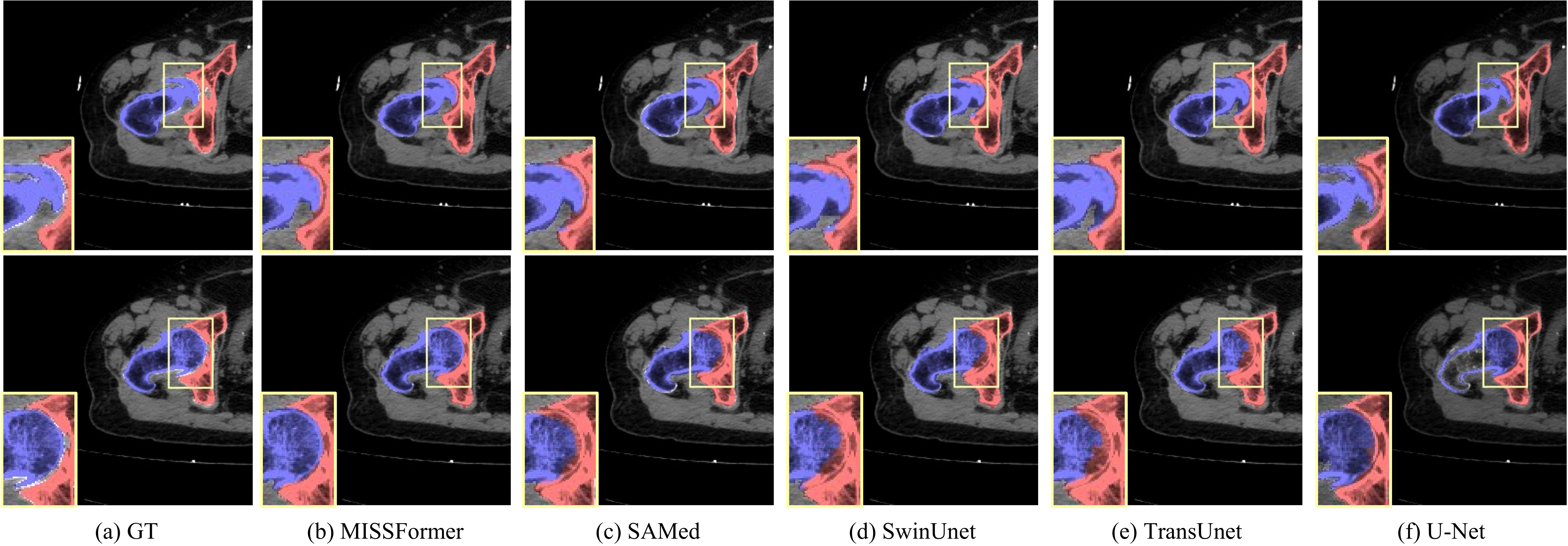}
\end{center}
   \caption{The qualitative comparisons between MISSFormer~\cite{9994763}, SAMed~\cite{samed}, SwinUnet~\cite{swinunet}, TransUnet~\cite{chen2021transunet} and U-Net~\cite{ronneberger2015u} on 2D CT slices. We illustrate two cases in different rows.}
\label{fig:inference_2d}
\end{figure*}

\begin{figure*}[t]
\begin{center}
\includegraphics[width=1.0\linewidth]{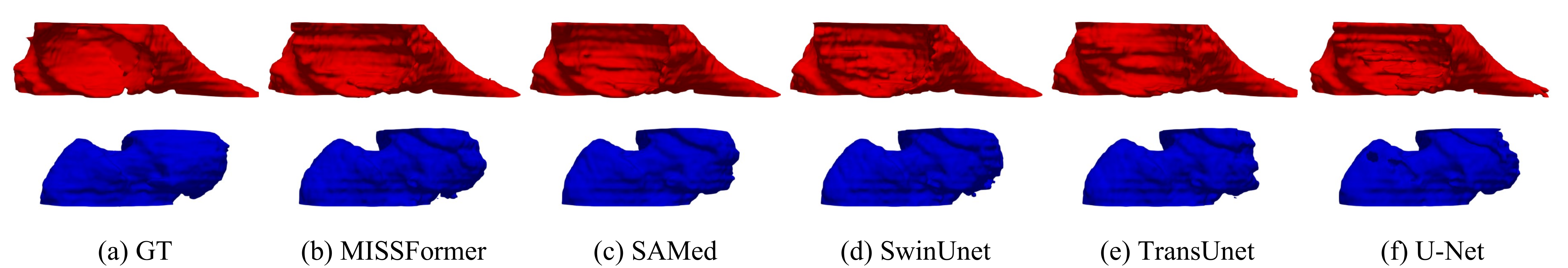}
\end{center}
   \caption{The rendered 3D results between MISSFormer~\cite{9994763}, SAMed~\cite{samed}, SwinUnet~\cite{swinunet}, TransUnet~\cite{chen2021transunet} and U-Net~\cite{ronneberger2015u}. The top and bottom row illustrate the acetabulum (red) and femoral head (blue), respectively. All the 3D scans are shown in the same viewpoint.}
\label{fig:inference_3d}
\end{figure*}

\subsubsection{Acetabulum and femoral head segmentation} We synthesize high quality acetabulofemoral masks with AFS and AAR to construct the dataset for finer bone structure analyses of THA. We provide the visualization of 2D and 3D bone structure of the acetabulum and the femoral head in Fig.~\ref{fig:train_2d} and Fig.~\ref{fig:train_3d}, respectively. The segmentation masks synthesized by our method exposes consistent patterns and high quality of the segmentation between acetabulum and femoral head. Although there is some room for improvement in terms of the accuracy of the femoral head, the quality of the synthesized segmentation mask is enough for the construction of a high quality dataset to provide valid examples for clinical usage and empower deep learning to perform bone structure analyses in finer granularity.

To conduct objective evaluation of the segmentation between acetabulum and femoral head, we annotate the testset manually under the supervision and correction from the experts in the hospital. We illustrate the manually annotated axial segmentation masks in Fig.~\ref{fig:test_2d} and the corresponding 3D bone structure of acetabulum and femoral head in Fig.~\ref{fig:test_3d}. Thanks to the professional knowledge from the experts, we successfully capture the accurate boundary of the femoral head. We believe the accurate annotation of the testset can provide precise measurement of the segmentation between acetabulum and femoral head.

For the construction of the benchmark, we train and evaluate multiple deep learning models on our proposed dataset for the segmentation between femoral head and acetabulum. We present the qualitative comparisons between five methods, including one convolution-based method U-Net~\cite{ronneberger2015u}, three transformer-based method TransUnet~\cite{chen2021transunet}, SwinUnet~\cite{swinunet} and MISSFormer~\cite{9994763}, and one PEFT-based method SAMed~\cite{samed}. Fig.~\ref{fig:inference_2d} and Fig.~\ref{fig:inference_3d} depict the generated segmentation results in 2D and 3D manners. All the methods are capable of achieving reasonable results based on our synthesized labels, which demonstrates the validity of our proposed annotation synthesis pipeline. The synthesized labels expose consistent and stable patterns about acetabulum and femoral head, which is crucial for the successful convergence of the deep learning models. Due to the difference of model capacity, various models achieve diverse outcomes. In the larger context, MISSFormer~\cite{9994763} can capture more accurate boundary between acetabulum and femoral head. What's more, the 3D structure of MISSFormer also maintains better contour of different bone structures. Our dataset can benefit the training of various deep learning structure (convolution and transformer) and training paradigm (PEFT-based method), which exhibits the versatility of our dataset.

\subsection{Ablation studies} In this section, we perform all the experiments based on three representative medical image segmentation methods - TransUnet~\cite{chen2021transunet}, SwinUnet~\cite{swinunet} and SAMed~\cite{samed}.

\begin{table}[t!]
\fontsize{8}{9}\selectfont
\centering
\caption{The comparisons of the segmentation accuracy of the models trained on the dataset with or without annotation filling}
% \footnotesize
% \resizebox{\textwidth}{!}{
\setlength{\tabcolsep}{1mm}{
\begin{tabular}{ccccccc}
\toprule
\multirow{2}{*}{Annotation filling} & \multicolumn{2}{c}{TransUnet} & \multicolumn{2}{c}{SwinUnet} & \multicolumn{2}{c}{SAMed} \\ \cmidrule(l){2-3} \cmidrule(l){4-5} \cmidrule(l){6-7}
& DSC$\uparrow$ & HD$\downarrow$ & DSC$\uparrow$ & HD$\downarrow$ & DSC$\uparrow$ & HD$\downarrow$ \\ \midrule
- & 81.70 & 8.90 & 81.83 & 7.94 & 80.97 & 8.47 \\
\checkmark & \textbf{89.90} & \textbf{8.85} & \textbf{90.09} & \textbf{7.67} & \textbf{88.82} & \textbf{8.30} \\
\bottomrule
\end{tabular}}
% }
\label{filling}
\end{table}

\begin{figure}[t]
\begin{center}
\includegraphics[width=0.6\linewidth]{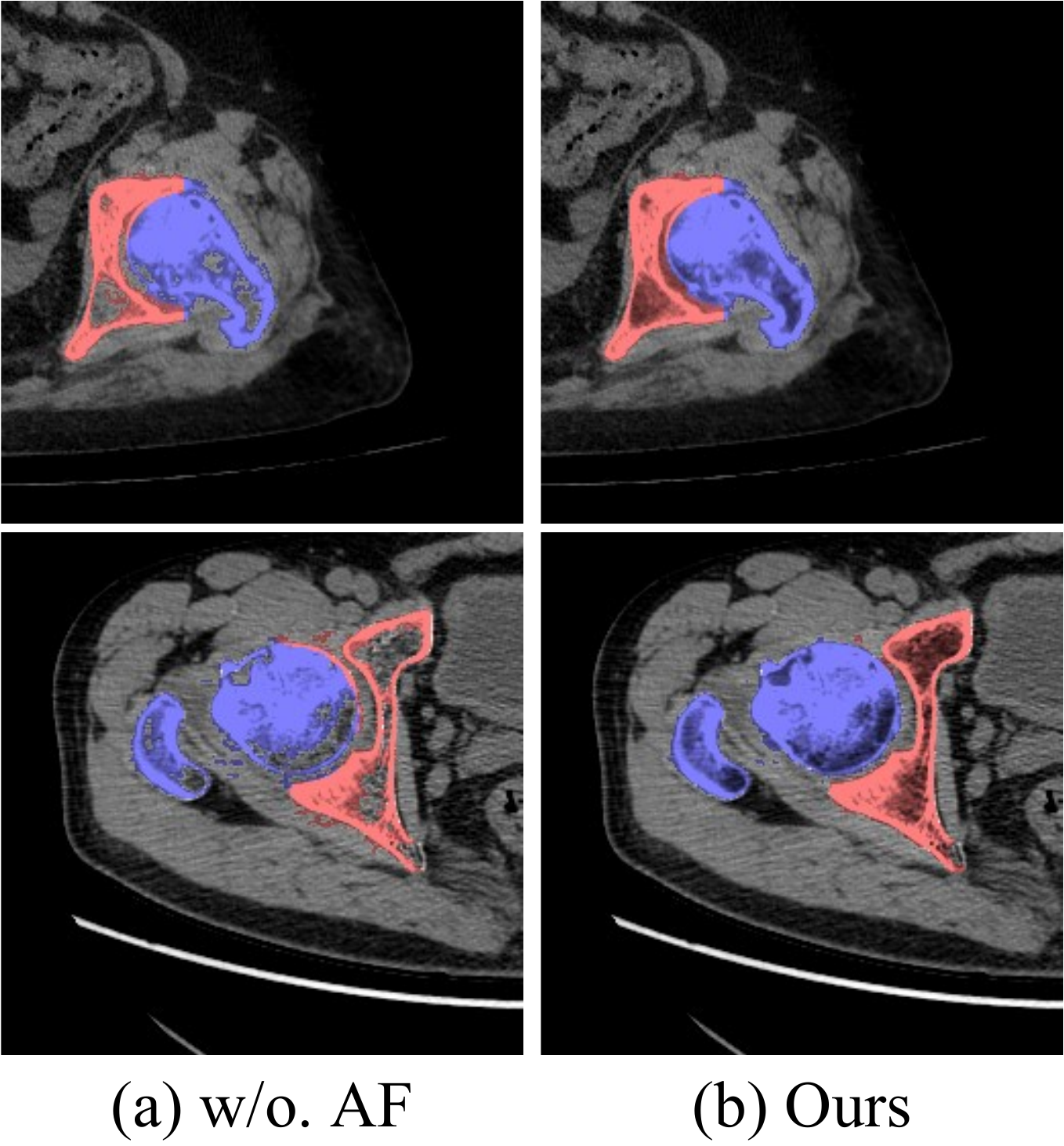}
\end{center}
   \caption{The comparisons of the results with and without annotation filling. We illustrate both the results from the training set (the top row) and the results generated by TransUnet~\cite{chen2021transunet} (the bottom row).}
\label{fig:AF}
\end{figure}

\subsubsection{Data preprocessing} To maintain the continuity of the bone tissues, we fill the binary holes in the segmentation masks. The binary holes left in the annotations harms the continuity of the predictions of the trained deep learning models, which leads to sub-optimal performance. We illustrate the performance comparisons in Tab.~\ref{filling}. The continuity of the annotations is essential for the improvement of dice score, which demonstrate the importance of the annotation filling procedure.

We illustrate the impact of annotation filling to training and test set in Fig.~\ref{fig:AF}. We present the samples from training set in the top row and the segmentation results generated by TransUnet~\cite{chen2021transunet} trained on our dataset in the bottom row. We note that the training set without annotation filling fails at maintaining the continuity of the bone tissues although they can still capture the boundary between acetabulum and femoral head. And such artifact in the training set deteriorates the performance of the corresponding trained deep learning models. Therefore, the construction of a high fidelity dataset is crucial for the further utilization. With annotation filling, both the training and test set capture the bone tissues with higher fidelity, which is also beneficial for the improvement of the model predictions.

\begin{table}[t!]
\fontsize{8}{9}\selectfont
\centering
\caption{Ablation study on the effects of our innovations in AFS}
% \footnotesize
% \resizebox{\textwidth}{!}{
\setlength{\tabcolsep}{1mm}{
\begin{tabular}{ccccccccc}
\toprule
\multirow{2}{*}{FSG} & \multirow{2}{*}{L-NMS} & \multirow{2}{*}{AHS} & \multicolumn{2}{c}{TransUnet} & \multicolumn{2}{c}{SwinUnet} & \multicolumn{2}{c}{SAMed} \\ \cmidrule(l){4-5} \cmidrule(l){6-7} \cmidrule(l){8-9}
& & & DSC$\uparrow$ & HD$\downarrow$ & DSC$\uparrow$ & HD$\downarrow$ & DSC$\uparrow$ & HD$\downarrow$ \\ \midrule
- & - & - & 78.46 & 14.93 & 81.61 & 11.61 & 81.47 & 11.43 \\
\checkmark & - & - & 83.66 & 11.00 & 84.95 & 10.67 & 84.16 & 10.08 \\
\checkmark & \checkmark & - & 88.39 & \textbf{8.78} & 89.34 & \textbf{7.60} & 88.58 & \textbf{7.33} \\
\checkmark & \checkmark & \checkmark & \textbf{89.90} & 8.85 & \textbf{90.09} & 7.67 & \textbf{88.82} & 8.30 \\
\bottomrule
\end{tabular}}
% }
\label{SAF}
\end{table}

\begin{figure}[t]
\begin{center}
\includegraphics[width=1\linewidth]{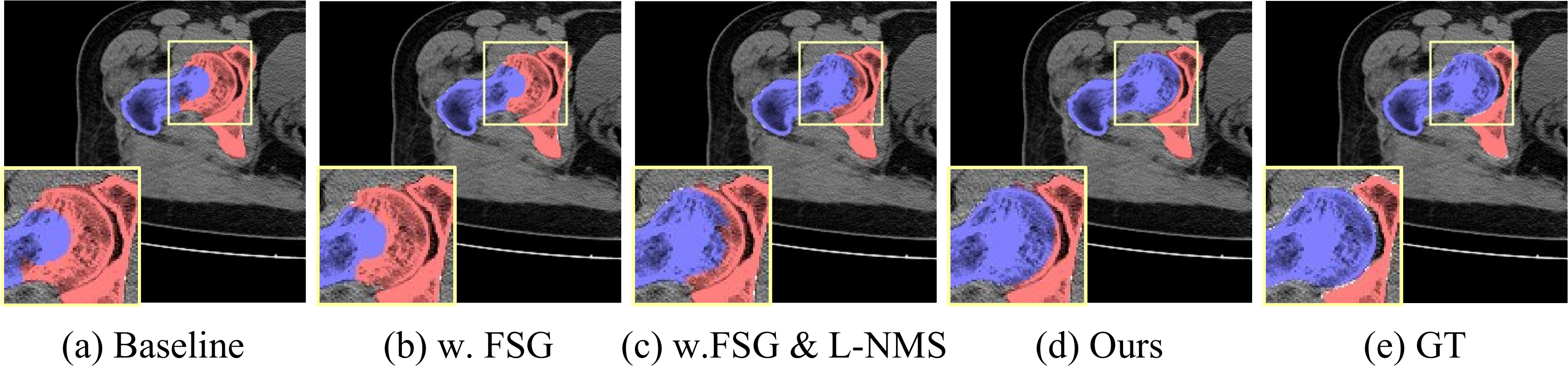}
\end{center}
   \caption{The qualitative comparisons of the effects between different proposed components in AFS. Baseline: the method without any innovation in AFS. Ours: the method with all of the innovations in AFS. The results are generated by TransUnet~\cite{chen2021transunet}.}
\label{fig:reg}
\end{figure}

\subsubsection{Effects of the proposed components in AFS} To synthesize more accurate annotations for femoral head and acetabulum, we propose three innovations to regularize the captured boundary in the femoral head, including first-order and second-order gradient regularization (FSG), line-based no-max suppression (L-NMS) and the anatomy-based Hough circle selection (AHS). These three components all improve the quality of annotations significantly. For ablation study, we gradually take of these three components in AFS and synthesize new annotations. Then, we train the deep learning models with the newly synthesized dataset to view the performance variance. 

As shown in the results from Tab.~\ref{SAF}, all of these three components lead to performance improvement of the trained deep learning models, which demonstrates the gradual refinement of the synthesized annotations. Annotations with higher fidelity expose more accurate patterns, which improves the generalization ability of the deep learning models significantly. We note that FSG and L-NMS are crucial for the high fidelity of the annotations, while the improvement of AHS is relatively small. We conjugate the confidence of each circle detected by Hough transform can instruct the selection of the boundaries of the femoral head to some extent, but such selection may be not as accurate as the anatomy prior in the hard examples. 

Fig.~\ref{fig:reg} illustrates the annotations synthesized by different variants of our method. The difference between these variants mainly lies in the accuracy of the captured boundaries between acetabulum and femoral head. With the introduction of these components, we observe stable improvement on the accuracy of the captured boundaries. Such results demonstrate the validity and effectiveness of the proposed innovations in AFS. With these innovations, we can construct more accurate dataset and train models with better performance for the segmentation between acetabulum and femoral head.

\begin{table}[t!]
\fontsize{8}{9}\selectfont
\centering
\caption{Ablation study on the effects of AAR for annotation refinement}
% \footnotesize
% \resizebox{\textwidth}{!}{
\setlength{\tabcolsep}{1mm}{
\begin{tabular}{ccccccc}
\toprule
\multirow{2}{*}{Method} & \multicolumn{2}{c}{TransUnet} & \multicolumn{2}{c}{SwinUnet} & \multicolumn{2}{c}{SAMed} \\ \cmidrule(l){2-3} \cmidrule(l){4-5} \cmidrule(l){6-7}
& DSC$\uparrow$ & HD$\downarrow$ & DSC$\uparrow$ & HD$\downarrow$ & DSC$\uparrow$ & HD$\downarrow$ \\ \midrule
baseline & 89.90 & 8.85 & 90.09 & 7.67 & 88.82 & 8.30 \\
random & 90.09 & 7.96 & 90.18 & 7.57 & 88.55 & 8.46  \\
AAR & \textbf{90.15} & \textbf{7.45} & \textbf{90.43} & \textbf{7.41} & \textbf{89.08} & \textbf{8.12} \\
\bottomrule
\end{tabular}}
% }
\label{AAR}
\end{table}

\subsubsection{Effects of AAR in annotation refinement} We adopt active learning to refine the annotations from the perspective of statistics. With active learning, we can easily locate and reannotate the samples with higher uncertainty. To demonstrate the effectiveness of annotation refinement, we compare our method with (a) the deep learning models trained without AAR, and we denote it as the ``baseline"; (b) the deep learning models trained on the annotations with manual reannotation to random samples (random).  We report the results in Tab.~\ref{AAR}. Compared with baseline, the improvement of manual annotation in both random sampling and AAR is restricted to a small range, which means the original annotations synthesized by our proposed annotation synthesis pipeline could achieve remarkable performance, therefore the left space for further improvement by the manual annotation is relatively small. Moreover, AAR achieves the best performance among all the methods and metrics, which validates the effectiveness of active learning in localizing the hard examples. And the targeted reannotation leads to the quality enhancement of the dataset.

% We illustrate the annotations before / after we adopt AAR for annotation refinement in Fig.9. We observe that AAR tends to select the samples with severe lesions and the upper or lower boundary of the femoral head for reannotation. These samples usually does not fit the anatomy prior in the acetafemoral regions, and reannotation for these samples leads to greater benefits for the overall annotation quality.

% To save labor cost, we only reannotate about 200 samples, which only occupies a very small fraction of the entire dataset. In Fig.10, we discuss the variance segmentation accuracy w.r.t. the number of selected samples in AAR. We identify the increment of the number of the selected samples lead to the improvement of the performance, but the improvement shrinks gradually. If we annotate more samples manually, we may achieve higher results, but the label cost will increase dramatically, too. Therefore, the strategy to reannotate 200 samples is reasonable in practice.

\section{Conclusion} This work presents a dataset for deep learning-based bone structure analyses in total hip arthroplasty and an annotation synthesis pipeline for automatic annotation with remarkable annotation quality and marginal labor and time costs. This pipeline includes bone extraction (BE), acetabulum and femoral head segmentation (AFS) and active learning based annotation refinement (AAR). In BE, we apply the graph-cut algorithm to extract the bone structures of the CT scans. In AFS, to improve the quality of the synthesized annotations, we design the first-order and second-order gradient regularization and the line-based no-max suppression to filter out the false positive points in the boundary of femoral head. Moreover, we propose to utilize the calculated anatomy circle to select more accurate boundaries of the femoral head. In AAR, we utilize the statistical knowledge and structural prior in the deep learning models to refine the annotations. We conduct extensive experiments to demonstrate the effectiveness of these components in synthesizing more accurate and comprehensive annotations. With our proposed annotation generation method, we construct a large-scale dataset for clinical analyses and deep learning in THA. The dataset is large-scale, clinical and diverse. We benckmark this dataset with multiple current mainstream medical image segmentation methods. 

This paper provides a novel perspective for data-centric medical AI. With the prior from medical knowledge, it's feasible to synthesize the medical annotations. The core of this problem is how to improve the annotation quality with limited labelling cost, and this problem is far not solved. Although our annotation generation pipeline achieves remarkable annotation synthesis ability, there is still a gap between our method and the manual annotations. In the future, we will  explore how to improve the annotation synthesis ability of our method and investigate the usage of the anatomy prior of other organs to construct high quality and large-scale dataset with marginal labelling cost.

% \section{CRediT author statement}
% \textbf{Kaidong Zhang}: Methodology, Software, Validation, Formal analysis, Investigation, Data Curation, Writing - Original Draft, Visualization.
% \textbf{Ziyang Gan}: Methodology, Validation, Investigation, Resources, Data Curation, Writing - Review \& Editing.
% \textbf{Dong Liu}: Conceptualization, Formal analysis, Resources, Writing - Review \& Editing, Supervision, Project administration, Funding acquisition.
% \textbf{Xifu Shang}: Conceptualization, Resources, Supervision.

\section{Acknowledgment}
This work was supported by the Fundamental Research Funds for the Central Universities under Contract WK3490000006.

%%Harvard
\bibliographystyle{model2-names.bst}\biboptions{authoryear}
\bibliography{refs}

% \section*{Supplementary Material}

% Supplementary material that may be helpful in the review process should
% be prepared and provided as a separate electronic file. That file can
% then be transformed into PDF format and submitted along with the
% manuscript and graphic files to the appropriate editorial office.

\end{document}